\documentclass[aps,pra,twocolumn,groupedaddress,superscriptaddress,bibnotes,amsfonts,citeautoscript,a4paper]{revtex4-1}
\usepackage{latexsym}
\usepackage{amsmath}
\usepackage{amssymb}
\usepackage{graphicx}
\usepackage{epstopdf}
\usepackage[T1]{fontenc}
\usepackage[open]{bookmark}
\usepackage{hyperref}
\hypersetup{colorlinks=true,allcolors=blue}
\usepackage{orcidlink}
\usepackage{epsfig}
\usepackage{epstopdf}
\usepackage{soul}
\usepackage{float}
\usepackage{ragged2e}
\usepackage{amsfonts}
\usepackage{color}
\usepackage{dsfont}
\usepackage[percent]{overpic}
\usepackage{lineno}
\usepackage[normalem]{ulem}
\usepackage{mathrsfs}

\begin{document}

\title{Dirac plasmon polaritons and magnetic modes in topological-insulator nanoparticles}

\author{Nikolaos Kyvelos\,\orcidlink{0000-0002-5164-1479}}
\email{niky@mci.sdu.dk}
\affiliation{POLIMA---Center for Polariton-driven Light--Matter Interactions, University of Southern Denmark, Campusvej 55, DK-5230 Odense M, Denmark}
\author{Vassilios Yannopapas\,\orcidlink{0000-0002-7832-7527}}
\affiliation{Department of Physics, School of Applied Mathematical and Physical Sciences, National Technical University of Athens, GR-15780 Athens, Greece}
\author{N. Asger Mortensen\,\orcidlink{0000-0001-7936-6264}}
\affiliation{POLIMA---Center for Polariton-driven Light--Matter Interactions, University of Southern Denmark, Campusvej 55, DK-5230 Odense M, Denmark}
\affiliation{Danish Institute for Advanced Study, University of Southern Denmark,
Campusvej 55, DK-5230 Odense M, Denmark}
\author{Christos Tserkezis\,\orcidlink{0000-0002-2075-9036}}
\email{ct@mci.sdu.dk}
\affiliation{POLIMA---Center for Polariton-driven Light--Matter Interactions, University of Southern Denmark, Campusvej 55, DK-5230 Odense M, Denmark}

\date{\today}

\begin{abstract}
We demonstrate the existence of previously unreported magnetic
modes with record-high magnetic Purcell factors in topological-insulator
nanospheres. Focusing on bismuth selenide (Bi$_{2}$Se$_{3}$), and based
on full electromagnetic Mie theory, we find magnetic modes arising from
both displacement current loops in the bulk, and surface
currents due to delocalized surface states, induced by electronic
transitions between topologically protected states within the Dirac
cone and discretized due to the sphere finite size.
Furthermore, we discuss how Dirac plasmon polaritons, resulting
from the interaction between THz photons and Dirac electrons,
dramatically influence both the magnetic and the electric transitions
of quantum emitters placed near Bi$_2$Se$_3$ nanospheres, significantly
enhancing the corresponding Purcell factors. These findings 
position Bi$_{2}$Se$_{3}$ nanospheres, whose optical response
is related to a richness of physical mechanisms, among the most
promising candidates for enhancing light--matter interactions in
nanophotonics and THz technologies.

\end{abstract}

\maketitle

\section{Introduction}\label{sec:intro}
Achieving strong coupling between light and atoms is one of the main
focuses of quantum optics~\cite{bernardot_epl17}, 
and has inspired nanophotonics to enter a quest for
confining electromagnetic (EM) modes within the smallest possible
volume, aiming to enhance light--matter interactions in the visible
and infrared~\cite{torma_rpp78,tserkezis_rpp83}.
To this end, plasmonics, with its tremendous field enhancement
and confinement, offers a natural template~\cite{fofang_nl8,zengin_prl114,
goncalves_natcom11,cuartero_acsphot5,kupresak_nscadv10,yuen-zhou_jcp156}. 
However, the inherent Ohmic losses in noble metals~\cite{boriskina_aop9,
khurgin_natnano10} tend to hinder their exploitation in quantum
technologies, and efforts have been made to identify alternatives,
e.g. in high-index dielectrics~\cite{baranov_lpr11,tserkezis_prb98}.
Dielectrics host EM states (Mie resonances) of both electric and
magnetic character which, on some occasions, dominate the frequency
window of interest~\cite{etxarri_oex19,evlyukhin_nl12}, and provide
flexibility in tailoring the decay rate of quantum emitters 
(QEs)~\cite{stamatopoulou_osac4}. Therefore, they offer the possibility
of controlling magnetic transitions of atoms via the magnetic Purcell
effect~\cite{schmidt_oex13,brule_oex30}. Although the Purcell effect
was originally associated with magnetic transition in atoms~\cite{purcell_pr69},
most studies so far, both with plasmonic and dielectric photonic
environments, focus on electric-dipole QEs; ways of achieving
strong light--matter interactions with magnetic-dipole QEs are
still limited~\cite{sloan_prb100}.

Topological insulators (TIs)~\cite{moore_nat464} have emerged
in the last couple of decades as prominent candidates for applications
in both quantum technologies~\cite{kitaev_prl96,qi_prl108,tschernig_natcom12}
and topological photonics~\cite{lu_natphot8,mortensen_nanoph8,
rider_acsphot9}. 
The latter area is practically divided into two subfields dominated by
different physics. One is related to the exploration of photonic EM
states endowed with topological protection~\cite{hafezi_natphot7,barik_sci359},
as observed in photonic-crystal architectures inspired by solid-state
systems~\cite{yannopapas_prb84,christensen_prx12,tang_lpr16}, mainly as
a result of time-reversal symmetry~\cite{wang_nat461,khanikaev_natmat12,
rosiek_natphot17}.
The other involves electronic topological systems interacting with 
light~\cite{dipietro_natnano8,ginley_aom6}. In this work, we focus
solely on the second situation, where the topological phenomena are
embodied in the electronic structure of the host material. Specifically,
we theoretically investigate the manipulation of Dirac plasmon polaritons
(DPPs)~\cite{dipietro_prl124,chen_natcom13,ginley_mrscom8} hosted within
Bi$_{2}$Se$_{3}$ spherical nanoparticles (NPs), with the aim of
controlling the spontaneous emission of QEs. In this context, DPPs
represent EM eigenstates that arise from the collective coupling of
massless Dirac electrons~\cite{haldane_prl61} residing on the surface
of a TI.

Apart from DPPs, further opportunities for exploitation
of TIs in photonics emerge when molding them into nanostructures.
In a previous work~\cite{imura_prb86}, an in-depth analysis on
the electronic spectrum of a spherical TI of 
nanometer dimensions revealed
quantization of the surface states within the Dirac cone of the
electronic band structure. This finding was subsequently analyzed
in terms of its influence on the NP optics~\cite{siroki_natcom7},
where transitions between these discrete levels were shown to induce
a surface charge density manifesting in the optical spectra as what
was termed a surface topological particle (SToP) mode. Nevertheless,
this response has only been studied so far within the quasistatic
regime~\cite{rider_nscale12}. Here, we seek to identify what effects
are ignored in such a description. In particular, by employing fully
electrodynamic Mie-theory-based calculations, we ask what happens
when one considers the dynamics of this surface charge density, and
whether the resulting surface current can lead to magnetic optical
response. At the same time, since TIs are usually characterized by
very large permittivities, we investigate the possible emergence
of magnetic modes as a result of displacement currents, in
analogy with traditional Mie-resonant semiconductors such as
silicon~\cite{staude_nn7,kivshar_nl22}. Focusing on Bi$_{2}$Se$_{3}$
NPs, we demonstrate previously unreported EM states of magnetic
nature spanning over a wide window in the THz regime, and show
that these states are capable of strongly affecting atomic
transitions over a wide range of NP--atom distances and frequencies.
The magnetic Purcell factors (PFs) predicted in this work are among
the highest reported in literature, potentially enabling the control
of otherwise forbidden transitions~\cite{rivera_sci353}. We anticipate
that our work will trigger a further quest for artificial atoms with
transitions in the THz regime~\cite{zhang_ol35,forati_prb90}, or for
ways to engineer TI NPs so as to push their optical response towards
the visible.

\section{Methodology}\label{Sec:method}

\begin{figure}[ht]
\centering
\includegraphics[width=0.8\columnwidth]{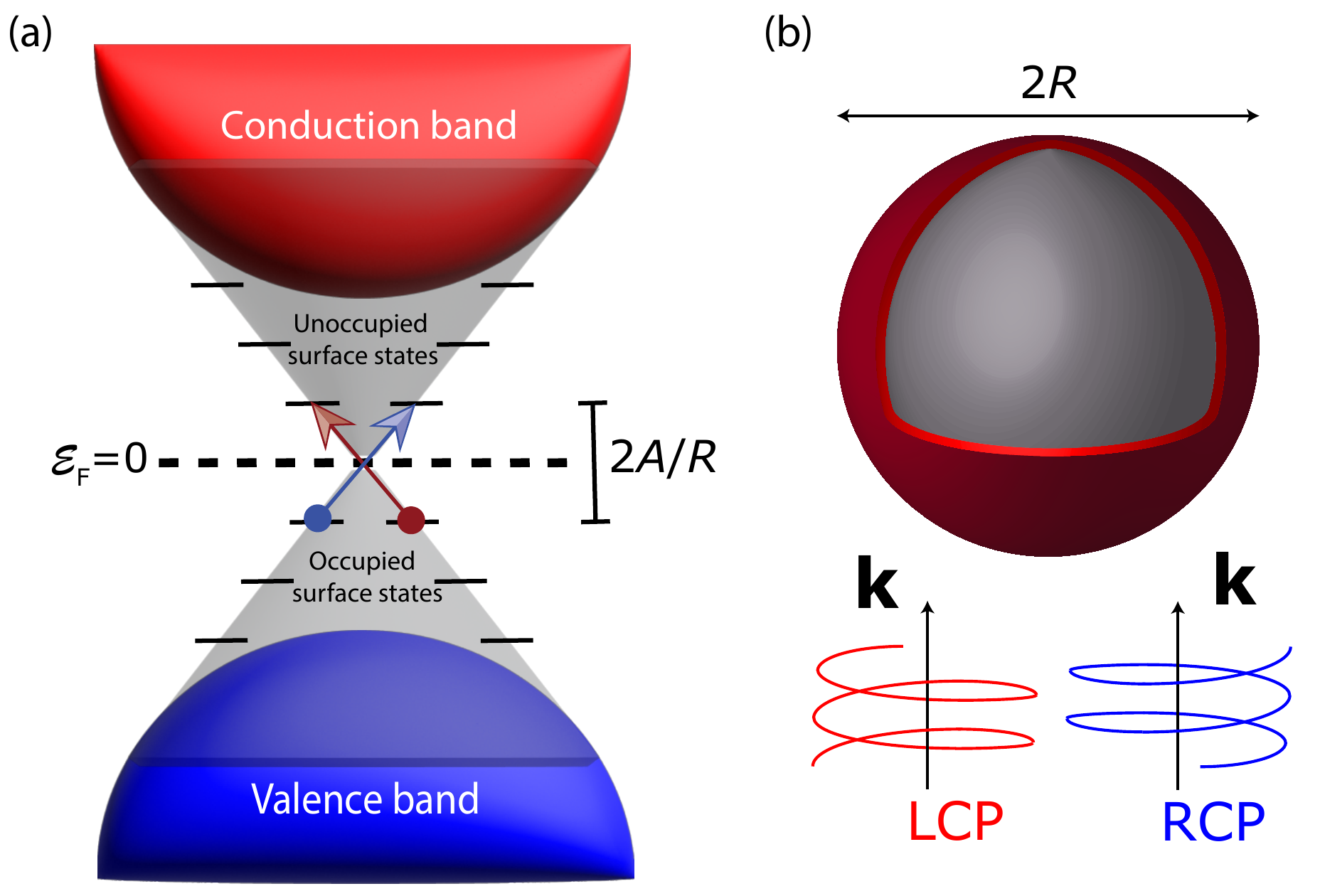}
\caption{
(a) Sketch of the band structure of a TI, with a band gap between the
valence and conduction bands, and a Dirac cone corresponding to surface
states. In small NPs, the surface Dirac cone exhibits discrete energy
levels, with energy spacing $\Delta \mathscr{E} = A/R$, where $A$ quantifies
the spin--orbit coupling strength. We assume that the Fermi energy 
$\mathscr{E}_{\mathrm{F}}$ is at the Dirac point. The color arrows
indicate allowed transitions between occupied and unoccupied surface
states.
(b) A spherical Bi$_{2}$Se$_{3}$ NP of radius $R$, excited by two
oppositely-handed circularly-polarized plane waves of wavevector
$\mathbf{k}$. Transitions between occupied and unoccupied states
lead to electronic surface states that manifest in the optics as
a surface current, sketched as an infinitesimally thin shell.
}
\label{fig1}
\end{figure}

The electronic configuration of TIs, in both two and three
dimensions, exhibits energy band gaps within the bulk, while featuring
gapless edge or surface states, usually characterized by a Dirac
dispersion~\cite{hasan_rmp82} [see sketch in Fig.~\ref{fig1}(a)].
Much like the case of graphene~\cite{kane_prl95b,jin_prl118,rappoport_acsphot6},
these topological surface states accommodate massless Dirac electrons
confined within a two-dimensional (2D) domain. The surface states
demonstrate strong spin--momentum coupling, resulting in remarkable
suppression of backscattering, as a direct consequence of time-reversal 
symmetry~\cite{moore_prb75}; electrons cannot backscatter between
distinct surface states unless there is a concurrent spin-flip event.
The outcome of this behavior is the emergence of unidirectional
surface currents~\cite{kastl_natcom6,luo_natcom10}.
In a three-dimensional (3D) TI~\cite{fu_prl98,hasan_arcmp2}, the surface
states are characterized by an odd number of Dirac cones---in the simplest
scenario a single one~\cite{kane_prl95a}. Among TIs, Bi$_{2}$Se$_{3}$
stands out for its relatively high bulk band gap of $0.3$\,eV, featuring
a single Dirac cone within its surface states. This substantial band gap
is attributed to the robust spin--orbit coupling, arising from the presence
of the heavy bismuth element~\cite{zhang_natphys5}. Upon reduction to
nanoscale structures, the escalating surface-to-volume ratio intuitively
leads to the anticipation that surface-related phenomena will significantly
impact the electrical and optical characteristics of the material. 
Spin--momentum locking~\cite{imura_prb84} of electrons within the TI
surface states opens possibilities for applications in THz 
sensing~\cite{zhang_prb82,tang_afm28}, spintronics~\cite{pesin_natmat11}
and for distinct phenomena such as spin-polarized plasmon
waves~\cite{kung_prl119,salikhov_natphys19}.

As discussed above, the transition from bulk to a nanostructured
TI comes with a natural, yet significant implication: the surface states
become quantized, leading to discrete energy levels encompassed by the
Dirac cone of the electronic band structure~\cite{imura_prb86}. These
energy levels are equidistant, and located symmetrically around the zero
of the energy axis (Dirac point), at intervals of $\Delta \mathscr{E} = 
A/R$, where $R$ denotes the NP radius. The parameter $A$ represents the
spin--orbit coupling strength, with density-functional theory (DFT)
predicting a value of $0.3$\,eV\,nm~ in the case of
Bi$_{2}$Se$_{3}$~\cite{liu_prb82}. The resulting spacing of
surface states is of the order of THz, highligting thus the potential
of TI NPs to bridge the so-called THz gap~\cite{low_nn8,zhang_natphot11}
and bring photonic systems closer to electronic ones. In what follows,
we assume that all occupied energy states extend up to the Dirac point,
implying that transitions occur from the lower to the upper part of the
Dirac cone, as illustrated in Fig.~\ref{fig1}(a). The charge density
that corresponds to interband transitions in the Dirac cone can be
calculated within time-dependent perturbation theory~\cite{siroki_natcom7}
in the dipole approximation. To address the optics of an NP with such
surface charges, in a zero-order approach one can incorporate the
resulting surface charge density into the NP polarizability, via an
angular frequency ($\omega$)-dependent function $\delta_{R} (\omega)$,
given as
\begin{equation}\label{delta}
\delta_{R} (\omega) = 
\frac{e^{2}}{6 \pi \varepsilon_{0} R}
\Bigg(\frac{1}{2 \Delta \mathscr{E} - \hbar\omega} +
\frac{1}{2 \Delta \mathscr{E} + \hbar \omega}\Bigg)
.
\end{equation}
This term originates from transitions occurring between the delocalized
topologically-protected surface states, which are perturbed by two
oppositely-handed, left- and right-circularly polarized (LCP/RCP) waves,
as sketched in Fig.~\ref{fig1}(b). It should be noted that, formally,
this expression does not adhere to causality, as it lacks an imaginary
part, 
since it does not take into account the lifetime of the transitions
between surface states; towards the end of the paper, we will address this
issue in a qualitative but still phenomenological way.

Introducing Eq~\eqref{delta} into the quasistatic polarizability,
the absorption cross section of a small nanosphere becomes
\begin{equation}\label{eq:sigma}
\sigma_{\mathrm{abs}} (\omega) = 
4 \pi R^{3} \frac{2\pi}{\lambda}
\mathrm{Im}
\Bigg[\frac{\varepsilon_{1} + \delta_{R} (\omega) - 1}
{\varepsilon_{1} + \delta_{R} (\omega) + 2}\Bigg]
,
\end{equation}
where $\lambda$ is the wavelength of the incident light, $\varepsilon_{1}$
is the relative permittivity of bulk Bi$_{2}$Se$_{3}$,
and the NP environment has been assumed to be air. This is a reasonable
starting point for studying the optical response of a system when the NP
size is much smaller than the wavelength of the incident field,
i.e., when the quasistatic approximation applies. The denominator of this
expression implies that a singularity should occur when $\varepsilon_{1} +
\delta_{R} (\omega) + 2 = 0$ and, indeed, a sharp resonance attributed
to the SToP mode has been theoretically predicted~\cite{siroki_natcom7} and
experimentally confirmed~\cite{rider_nscale12}.

In what follows, we examine the system within a fully electrodynamic
picture, aiming to serve a dual purpose: on the one hand, since
it is known that high-refractive-index dielectrics sustain both electric
and magnetic Mie resonances, the latter caused by displacement current loops
in the bulk of the NP, we wish to explore whether these are relevant in
the case of Bi$_{2}$Se$_{3}$. On the other hand, in a dynamic picture,
the surface charge of Eq.~\eqref{delta} should be seen as a topological
surface current, possibly leading to magnetic response as well. 
In addition, when studying the interaction of QEs with photonic
resonators, higher-order multipoles are known to be of vital
importance~\cite{anger_prl96}. Going beyond the quasistatic approximation
enables thus to capture the full range of optical modes of the system,
and more accurately predict its coupling to QEs. We should stress here
that there are two levels of dipole approximations in such a study; one
concerning the possible transitions within the Dirac cone, which
predicts the surface charge density that we use throughout the
manuscript, and one concerning the optical response of the NP, once
such a charge is placed on its surface. The former is only related to
microscopic selection rules, and there is no obvious reason to depart
from it.

Based on Mie theory~\cite{Bohren_Wiley1983}, it is possible to calculate
the scattering and absorption cross sections of Bi$_{2}$Se$_{3}$ nanospheres
excited by a monochromatic plane wave. The NP is placed in vacuum, ensuring
that the topologically non-trivial bulk is interfaced by a topologically
trivial background. The modification in topological invariant at the
interface necessitates the closure of the band gap, while simultaneously
preserving the gap within the bulk of both media. This configuration leads
to the manifestation of a topological phase transition at the NP surface,
thereby revealing the emergence of conductive topological surface states.
In the multipolar expansions of the incident, internal, and scattered
fields in spherical waves, we use appropriate expansion coefficients
$a_{P\ell m}^{0}$, $a_{P\ell m}^{\mathrm{in}}$, and $a_{P\ell m}^{+}$,
respectively. Here, the subscripts $\ell$ and $m$ correspond to the usual
angular momentum indices, while the polarization index $P = E, H$ refers
to transverse electric (TE) or transverse magnetic (TM) polarization,
respectively.

The expansion coefficients are determined by application of the appropriate
boundary conditions at the NP surface. In our case, the usual continuity of
the tangential components ($\parallel$) of the electric ($\mathbf{E}$) and
magnetic ($\mathbf{H}$) field must be modified to account for the induced
surface current, denoted here as $\mathbf{K}$, that arises from electrons
in the Dirac cone. The new boundary conditions read
\begin{subequations}
\begin{align}
\left( \mathbf{E}_{\parallel \mathrm{out}}  -
\mathbf{E}_{\parallel \mathrm{in}} \right)_{R}
&=0 \label{prwth}, \\
\left( \mathbf{H}_{\parallel \mathrm{out}} -
\mathbf{H}_{\parallel \mathrm{in}} \right)_{R}
&= \mathbf{K} \label{current}
,
\end{align}
\end{subequations}
through which we can extract the electric ($a_{\ell}$) and
magnetic ($b_{\ell}$) Mie scattering coefficients~\cite{Bohren_Wiley1983},
defined through $a_{E\ell m}^{+} = a_{\ell} a_{E\ell m}^{0}$ and
$a_{H\ell m}^{+} = b_{\ell} a_{H\ell m}^{0}$, as
\begin{subequations}\label{Eq:Mie2D}
\begin{align}
a_{\ell} &= 
\frac{ j_{\ell} (x_{1}) \Psi_{\ell}^{'} (x_{2}) \varepsilon_{1} -
j_{\ell} (x_{2}) \Psi_{\ell}^{'} (x_{1}) \varepsilon_{2} +
g (\omega, x_{1}) \Psi_{\ell}^{'} (x_{2})}
{h_{\ell}^{+} (x_{2}) \Psi_{\ell}^{'} (x_{1}) \varepsilon_{2} -
j_{\ell} (x_{1}) \xi_{\ell}^{'} (x_{2}) \varepsilon_{1} -
g (\omega, x_{1}) \xi_{\ell}^{'} (x_{2})}, \\
b_{l} &=
\frac{j_{\ell} (x_{1}) \Psi_{\ell}^{'} (x_{2}) \mu_{1} -
j_{\ell} (x_{2}) \Psi_{\ell}^{'} (x_{1}) \mu_{2} +
c (\omega, x_{1}) j_{\ell} (x_2)}
{h_{\ell}^{+} (x_2) \Psi_{\ell}^{'} (x_{1}) \mu_{2} -
j_{\ell} (x_{1}) \xi_{\ell}^{'} (x_{2}) \mu_{1} -
c (\omega, x_{1}) h_{ell}^{+} (x_2)}
,
\end{align}
\end{subequations}
where $x_{j} = \frac{\omega}{c} \sqrt{\varepsilon_{j} \mu_{j} } R$
with $c$ being the speed of light in vacuum, while
$\varepsilon_{j}$ and $\mu_{j}$ are the relative permittivity and
permeability of the sphere ($j = 1$) and the host medium ($j = 2$),
respectively. The corrections $g$ and $c$ are given by
\begin{subequations}\label{Eq:Miecg}
\begin{align}
g (\omega, x_{1}) &= 
\frac{\mathrm{i} \sigma \Psi_{\ell}^{'} (x_{1})} 
{\varepsilon_{0} \omega R}
,
\\
c (\omega, x_{1}) &= 
\frac{\mathrm{i} x_{0}^{2} \sigma j_{\ell} (x_{1}) \mu_{1} \mu_{2}} 
{\varepsilon_{0} \omega R}
,
\end{align}
\end{subequations}
with the index $0$ corresponding to vacuum. We assume that the
permeabilities of all media are equal to unity. In addition, we
introduced the Riccati--Bessel functions  $\Psi_{\ell} (x) =
x j_{\ell} (x)$ and $\xi_{\ell} (x) = x h_{\ell}^{+} (x)$,
where $j_{\ell}$ and $h_{\ell}^{+}$ are the spherical Bessel,
and Hankel of first kind functions of order $\ell$, respectively.
The surface conductivity
\begin{equation}\label{eq:sigma}
\sigma =
-\frac{\mathrm{i} \omega \delta_{R} \varepsilon_{0} R}{2}
,
\end{equation}
to which the surface current density $\mathbf{K}$ corresponds,
is obtained from the continuity equation $\boldsymbol{\nabla} \mathbf{K} =
- \frac{\partial \rho }{\partial t}$ at the NP surface, using the 2D
divergence theorem, as the surface charge density
$\rho = -a_{E\ell m}^{\mathrm{in}} \delta_{R} \varepsilon_{0}$ resides
precisely on the surface.

The scattering, extinction, and absorption cross sections, normalized
to their geometric cross section $\pi R^{2}$, can be expressed in terms
of the Mie coefficients as
\begin{equation}\label{Eq:CrosecsSpherical}
\begin{split}
\sigma_{\mathrm{sc}} &= 
\frac{2}{\left(k R\right)^{2}} 
\sum_{\ell}
\left(2\ell + 1\right) 
\left(\left|{a_{\ell}}\right|^{2} + 
\left|{b_{\ell}} \right|^{2}\right)
,\\
\sigma_{\mathrm{ext}} &= 
- \frac{2}{\left( kR \right)^{2}} 
\sum_{\ell}
\left(2\ell + 1\right) 
\mathrm{Re} \left({a_{\ell}}  + {b_{\ell}} \right)
,\\
\sigma_{\mathrm{abs}} &= 
\sigma_{\mathrm{ext}} -
\sigma_{\mathrm{sc}}
.
\end{split}
\end{equation}

Finally, experimental data are used for the dielectric function of
Bi$_{2}$Se$_{3}$~\cite{butch_prb81}, conveniently represented by a sum
of three Lorentzians,
\begin{equation}\label{epsbulk}
\varepsilon_{1} (\omega) =
\sum_{i = \alpha, \beta, f} 
\frac{\omega_{\mathrm{p} i}^{2}}
{\omega_{0 i}^{2} - \omega^{2} - \mathrm{i} \omega\gamma_{i}}
,
\end{equation}
plotted in Fig.~\ref{fig2}(a). The subscripts $i$ appearing in the sum
in Eq.~\eqref{epsbulk} stem from contributions of $\alpha$ and $\beta$ phonons,
as well as from bulk free-charge carriers $f$ at low frequencies. The
corresponding parameters for the three terms appearing in Eq.~\eqref{epsbulk}
are taken as
$\hbar \omega_{\mathrm{p} \alpha} = 79.4$\,meV,
$\hbar \omega_{\mathrm{p} \beta} = 9.5$\,meV, 
$\hbar \omega_{\mathrm{p} f} = 47.6$\, meV, 
$\hbar \omega_{0 \alpha} = 8.3$\,meV, 
$\hbar \omega_{0 \beta} = 15.4$\,meV,
$\omega_{0 f} = 0$ (by assumption), 
$\hbar \gamma_{\alpha} = 0.6$\,meV, 
$\hbar \gamma_{\beta} = 0.3$\,meV, and 
$\hbar \gamma_{f} = 1.0$\,meV.
We note that, since Bi$_{2}$Se$_{3}$ is anisotropic, we have assumed
that the optical axis of the NP coincides with the direction of the
wavevector of the incident EM wave, as illustrated in Fig.~\ref{fig1}(b).

\section{Results and discussion}

\begin{figure}[ht]
\centering
\includegraphics[width=0.95\linewidth]{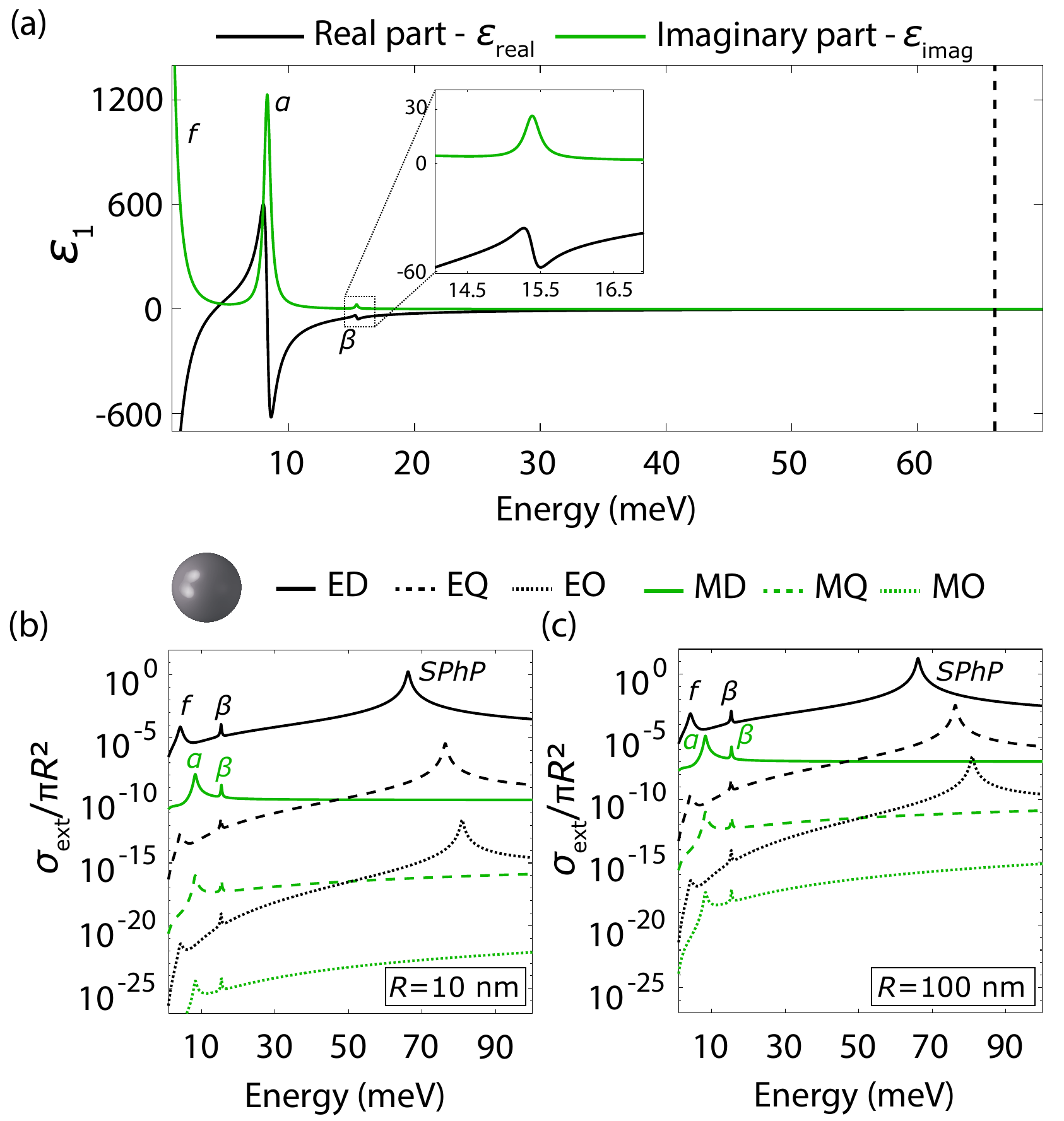}
\caption{(a) Real (black line) and imaginary part (green line) of
the dielectric function of bulk Bi$_{2}$Se$_{3}$~\cite{butch_prb81}
described by Eq.~\eqref{epsbulk}. The resonance around $8$\,meV corresponds
to the $\alpha$ phonon, while the smaller resonance at about $15$\,meV
(inset) corresponds to the $\beta$ phonon. The vertical dashed line
indicates the energy where $\mathrm{Re} \varepsilon_{1} = -2$. 
(b, c) Contributions of dipolar (ED/MD, solid lines), quadrupolar
(EQ/MQ, dashed lines), and octupolar (EO/MO, dotted lines) modes of
both electric (in black) and magnetic (in green) character to the
extinction cross section (log scale) of a Bi$_{2}$Se$_{3}$ nanosphere
described by the dielectric function of (a), and radii
(b) $R = 10$\,nm and
(c) $R =1 00$\,nm.
The resonances related to free electrons ($f$), the $\alpha$ and
$\beta$ phonons, and the SPhP, are marked in each panel.}
\label{fig2}
\end{figure}

In Figs.~\ref{fig2}(b) and (c) we show extinction spectra
in the absence of surface charges and currents, for two NP
sizes, $R = 10$\,nm and $R = 100$\,nm, normalized to the geometric cross
section $\pi R^{2}$, decomposed into electric/magnetic (black/green lines)
dipole (ED/MD: $\ell = 1$, solid lines), quadrupole (EQ/MQ: $\ell = 2$,
dashed lines), and octupole (EO/MO: $\ell = 3$, dotted lines) contributions.
It is evident that the ED contributions dominate the extinction spectrum
in both cases, in accordance with the quasistatic approximation
and previous studies of such NPs~\cite{siroki_natcom7}.
In addition, one can also see that MD modes related
to the high permittivities due to the $\alpha$ and $\beta$ phonons
are, in principle, present, albeit three to five orders of magnitude
weaker in these examples, and thus not contributing to the full spectrum;
nevertheless, their existence can become important when the NP interacts
with a QE, as we discuss later.
At low energies, around $4$\,meV, a plasmonic mode emerges, originating
from the $f$ term of the dielectric function associated with free electrons
within the bulk. These electrons arise from inevitable defects in the 
bulk~\cite{butch_prb81}. Subsequently, at about $15$\,meV, a resonance 
originating from the $\beta$ phonon coincides with the peak in the
dielectric function [see inset in Fig.~\ref{fig2}(a)], indicating
excitation of the $\beta$ \emph{bulk phonon polariton}. At even higher
energies, a \emph{surface phonon polariton} (\emph{SPhP}) becomes
prominent, emerging exactly where $\mathrm{Re} \varepsilon_{1} = -2$
[the Fr{\"o}hlich condition for a resonance within the quasistatic 
approximation according to Eq.~\eqref{eq:sigma}---for $\delta_{R} =0$];
the frequency at which this happens is indicated with a vertical dashed
line in Fig.~\ref{fig2}(a). This mode has already been proven
responsible for a strong anisotropic Purcell effect in the case of
microspheres~\cite{karaoulanis_josab38,kyvelos_phot9,papadaki_phot10}.
For a $10$\,nm radius [Fig.~\ref{fig2}(b)], the SPhP precisely aligns
with the $66$\,meV mark of Fig.~\ref{fig2}(a), while for larger radii
[$R = 100$\,nm in Fig.~\ref{fig2}(c)], it only shifts slightly, to
around $68$\,meV, due to retardation. In what follows, we
will demonstrate that taking into account the surface states introduces
DPPs, which lead to additional magnetic modes of topological nature,
dramatically altering the optical response of the system.

TIs are theoretically anticipated to manifest robust
conductivity on their surfaces, even in the presence of substantial
impurities~\cite{tokura_natrphys1}; in principle, TIs are perfect
conductors on their surfaces, potentially producing non-dissipative
surface currents. However, it is necessary to take into account the
coupling of the electrons in the Dirac cone with the bulk phonons,
which have a pronounced presence in Bi$_{2}$Se$_{3}$. Coupling of
an electron with a phonon results in the destruction of the surface
state and a topological phase change, i.e., the system transforms
into a trivial insulator, and the generation of surface current is
prevented. It should be emphasized that Bi$_{2}$Se$_{3}$ is a
room-temperature TI, meaning that even at a temperature of $300$\,K,
the Dirac cone persists, and the surface states are topologically
protected. In spherical NPs, it has been found~\cite{siroki_natcom7}
that for radii larger than $37.5$\,nm the surface states may be
disrupted due to interaction of electrons with phonons.
This means that, in what follows, results for radii up
to $37.5$\,nm can be considered valid at room temperature, while
for larger sizes one should assume a temperature of $6$\,K.

\begin{figure*}[ht]
\centering
\includegraphics[width=0.8\linewidth]{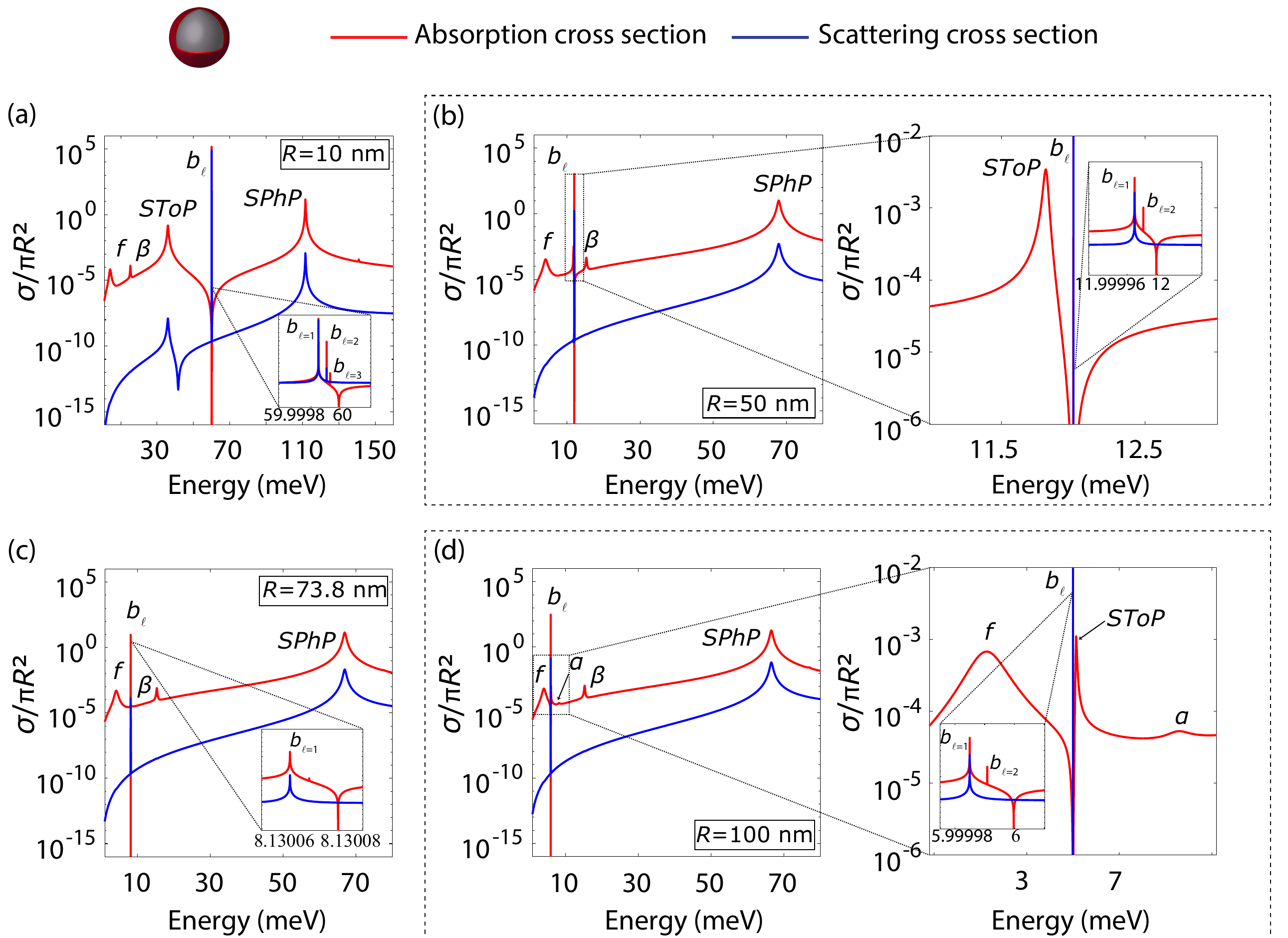}
\caption{Normalized absorption (red lines) and scattering (blue lines) 
cross sections (log scale) of Bi$_{2}$Se$_{3}$ nanospheres in which
transitions between discrete levels at the surface states have been
taken into account, for (a) $R = 10$\,nm, (b) $R = 50$\,nm,
(c) $R = 73.8$\,nm, and (d) $R = 100$\,nm. In panels (a) and (c),
the spectrum at the energy of the topological magnetic modes is
depicted in the insets, and analyzed into multipolar contributions.
In panel (a), the labels SToP and SphP characterize the hybrid modes
emerging from the interaction of the two.
In panel (c), the SToP mode is totally suppressed, as shown in Figure~S3.
In panels (b) and (d), the right-hand graphs are zooms into the frequency
window of the SToP and induced magnetic modes, with their insets further
focusing on just the magnetic multipoles. 
}\label{fig3}
\end{figure*}

For specific photon energies, the Bi$_{2}$Se$_{3}$ nanosphere, acting
as a TI, undergoes transitions between surface states in the Dirac cone.
In this work we consider transitions between the two lowest degenerate
states, with energy $\mathscr{E} = -A/R$, and the two highest degenerate
states, with energy $\mathscr{E} = A/R$, just below and above the Fermi
energy $\mathscr{E}_{\mathrm{F}} = 0$, which we take exactly at the Dirac
point [see Fig.~\ref{fig1}(a)]; this choice ensures an analytic expression
for the surface charge density that can directly be implemented as a
surface current in our calculations, while considering other transitions
would require a more numerical---and thus less transparent---treatment.
The response of the NP to an external electric field is predominantly
influenced by the electrons near the Fermi level, and the most probable
transition will occur between the aforementioned energy states. Importantly,
the response of surface states varies according to the polarization of
the incident radiation. This implies that a photocurrent will be generated
precisely on the surface of the sphere. The exponential decay of surface
states outside the surface region
(which can be estimated as approximately 0.5\,nm)~\cite{liu_prb82}
leads to the assumption that the topological surface current exists exactly
at $r = R$.

\begin{figure}[ht]
\centering
\includegraphics[width=0.9\linewidth]{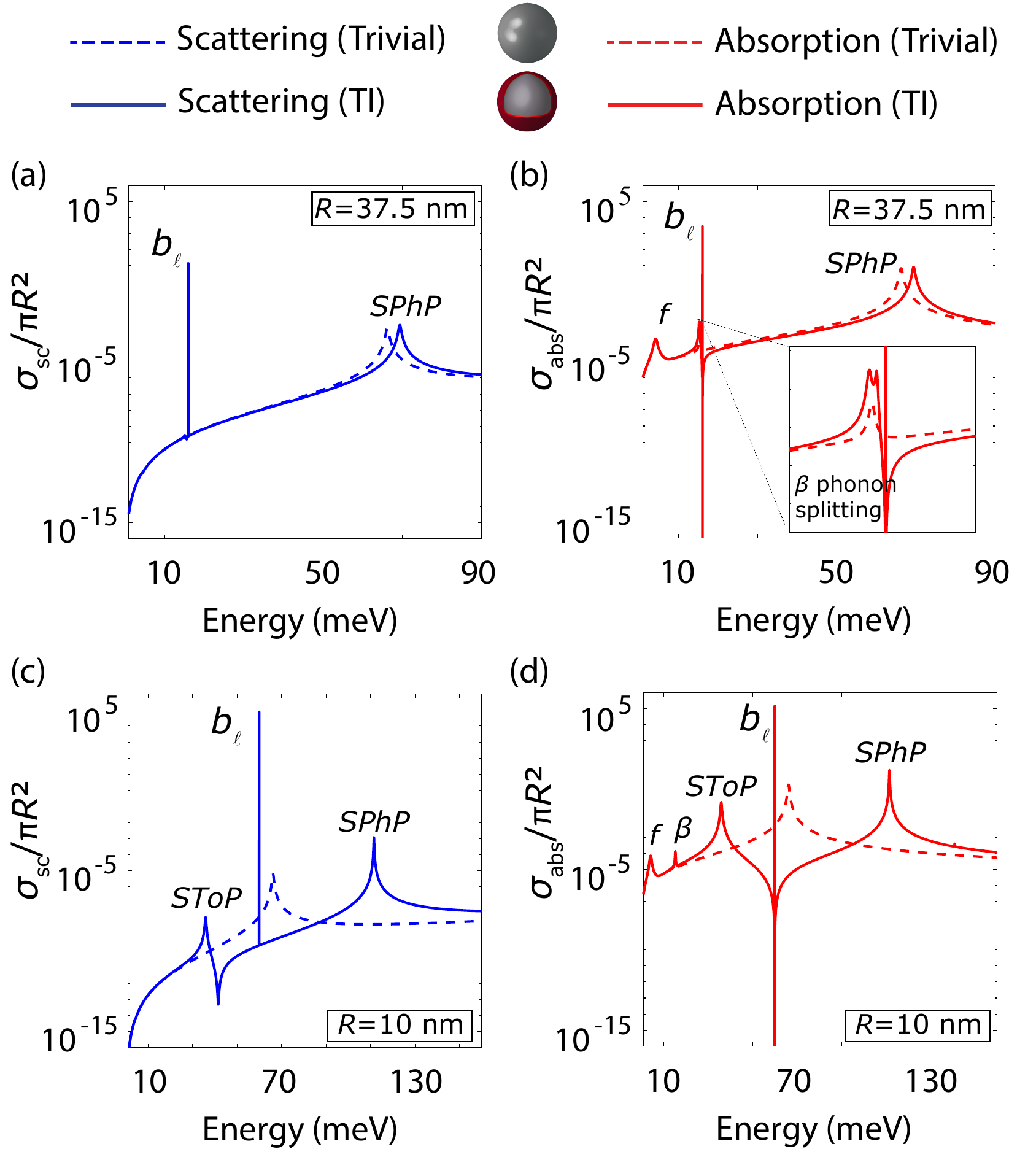}
\caption{
(a) Normalized scattering and (b) absorption cross section for a
trivial-insulator (dashed lines ) and a TI (solid lines)
Bi$_{2}$Se$_{3}$ nanosphere, (i.e, without and with surface states)
for $R = 37.5$\,nm. Splitting of the $\beta$ bulk phonon polariton
is shown in the zoom-in inset of (b).
(c), (d) Same as (a), (b), for $R= 10$\,nm. In this case, splitting
of the SPhP is observed in both scattering and absorption. The labels
SToP and SPhP refer to the lower- and higher-energy hybrid modes that
emerge from the interaction of the individual SToP and SPhP.}
\label{fig4}
\end{figure}

Absorption and scattering spectra for $R = 10$\,nm, $50$\,nm, $73.8$\,nm
(which is a special value, as we discuss below) and $100$\,nm are shown
in Fig.~\ref{fig3}, taking into account the surface states that give
rise to a SToP mode, affected by coupling to the $\alpha$ phonon
[for comparison with the quasistatic approximation, see also
Fig.~\ref{figS4}(a)]. It is established that interaction between the modes 
sustained by opposite surfaces can occur, resulting in the formation of
a gap in the Dirac cone. Experiments have shown that, in the case of Bi$_{2}$Se$_{3}$, a film thickness of less than six quintuple layers
is the limit at which a finite energy gap appears~\cite{zhang_natphys6}.
For this reason, the smallest radius examined below is set at $10$\,nm.
On the other hand, the radius
of $100$\,nm represents the maximum size that the nanosphere can attain
before transitioning to a nearly-continuum of surface states, with energy
differences outside our frequency window. The SToP mode, resulting
from the interaction of Dirac electrons with the $\alpha$ phonon, is
depicted in Fig.~\ref{fig3}(a) at about 36\,meV (this is, in
fact, a special case analyzed in more detail in Fig.~\ref{fig4}),
in Fig.~\ref{fig3}(b) at about $12$\,meV, and in (d) at about $6$\,meV,
and is of ED character ($a_{\ell = 1})$. For $R = 73.8$\,nm
[Fig.~\ref{fig3}(c)], the SToP mode disappears (see Fig.~\ref{figS2}), as it
is entirely absorbed by the $\alpha$ phonon~\cite{chatzidakis_prb101}.
In addition to 
these findings of previous studies, our fully electrodynamic
calculations further reveal---apart from the weak bulk magnetic modes
mentioned previously---the emergence of magnetic \emph{topological}
modes, denoted as $b_{\ell}$, manifesting as needle-like peaks in the
zoomed graphs; interestingly, these modes do survive for $R = 73.8$\,nm.
We also demonstrate the possibility of magnetic Mie modes stemming from
the bulk phonons (see Figure~S2) which can contribute to the extinction
cross sections in the case of large nanospheres, such as $100$\,nm,
as can be seen in Fig.~\ref{fig3}(d) ($\alpha$ phonon). The magnetic topological modes arise from the current at the surface of the TI.
Notably, a zero-absorption point can be observed for every radius,
where the electric field inside the NP tends to zero. This implies
that in this energy region the NP transforms into a perfect conductor,
allowing for the presence of surface currents originating from the
surface charge density generated from transitions between the discrete
levels in the Dirac cone with energy difference $\Delta\mathscr{E}$.

We also observe that by reducing the radius, the topological
modes are enhanced, and their linewidths are modified. As the NP decreases
in size, the ratio of bulk material to surface diminishes, resulting in
increased prominence of surface states. It is still clear that in each
case a point of zero absorption is detected, where the electric field
inside the NP is zero, and the nanosphere transforms into a perfect
conductor. Fig.~\ref{fig3}(d) shows that these topological
optical modes are essentially Fano resonances~\cite{limonov_natphot11},
as they have a narrow and asymmetric line-shape. Specifically, in the
case of the SToP mode, the discrete surface state excitation corresponds
to the resonant process, while the coupling with the phonon mode, which
is broader in comparison, corresponds to the background process. On the
other hand, there is no interaction of the magnetic modes $b_\ell$ with
a background process, resulting in the appearance of extremely sharp
resonances, with the linewidth again related to the absence of an 
imaginary part in Eq.~\eqref{delta}. The multipolar analysis of the
spectra is further extended in Fig.~\ref{figS1}, wherein a detailed
representation of the contributions from all multipoles in the
$R = 100$\,nm NP is depicted.

To shed more light on the influence of surface states on the spectra,
we examine various NP radii, which adjust the energy at which the
surface charge density emerges, as it depends on the factor $2A/R$.
In Fig.~\ref{fig4}, it becomes evident that the surface states
induce a shift of the polaritonic modes, implying a splitting of
the modes. The radii in the figure are chosen in such a way that
the surface states coincide with similar resonant frequencies of
the modes of the trivial-insulator NP. In the case of $R = 37.5$\,nm
[Figs.~\ref{fig4}(a) and (b)], when the surface states are excited in
the frequency region of the $\beta$ phonon, a splitting of the bulk
phonon polariton mode occurs in absorption, as can be seen in the
inset of Fig.~\ref{fig4}(b). However, this splitting is not pronounced,
as it involves interactions between the bulk optical mode and the surface
optical mode. On the other hand, in the case of $R = 10$\,nm we observe
a very strong splitting of the SPhP mode, in both scattering and
absorption, revealing a new situation of self-hybridization
between different states of the same system~\cite{canales_jcp154,
tserkezis_nanoph13}, as illustrated in Figs.~\ref{fig4}(c) and (d) at
$36$\,meV. This response can be interpreted as follows: the topological
surface current results in blocking the emergence of the SPhP, shifting
it to side frequencies where the current vanishes. This is clearly
visible by Figs.~\ref{fig4}(c) and (d) featuring the distinct magnetic
topological mode between resonances with a mixed SToP and SPhP
character.

\begin{figure}[ht]
\centering
\includegraphics[width=1.0\linewidth]{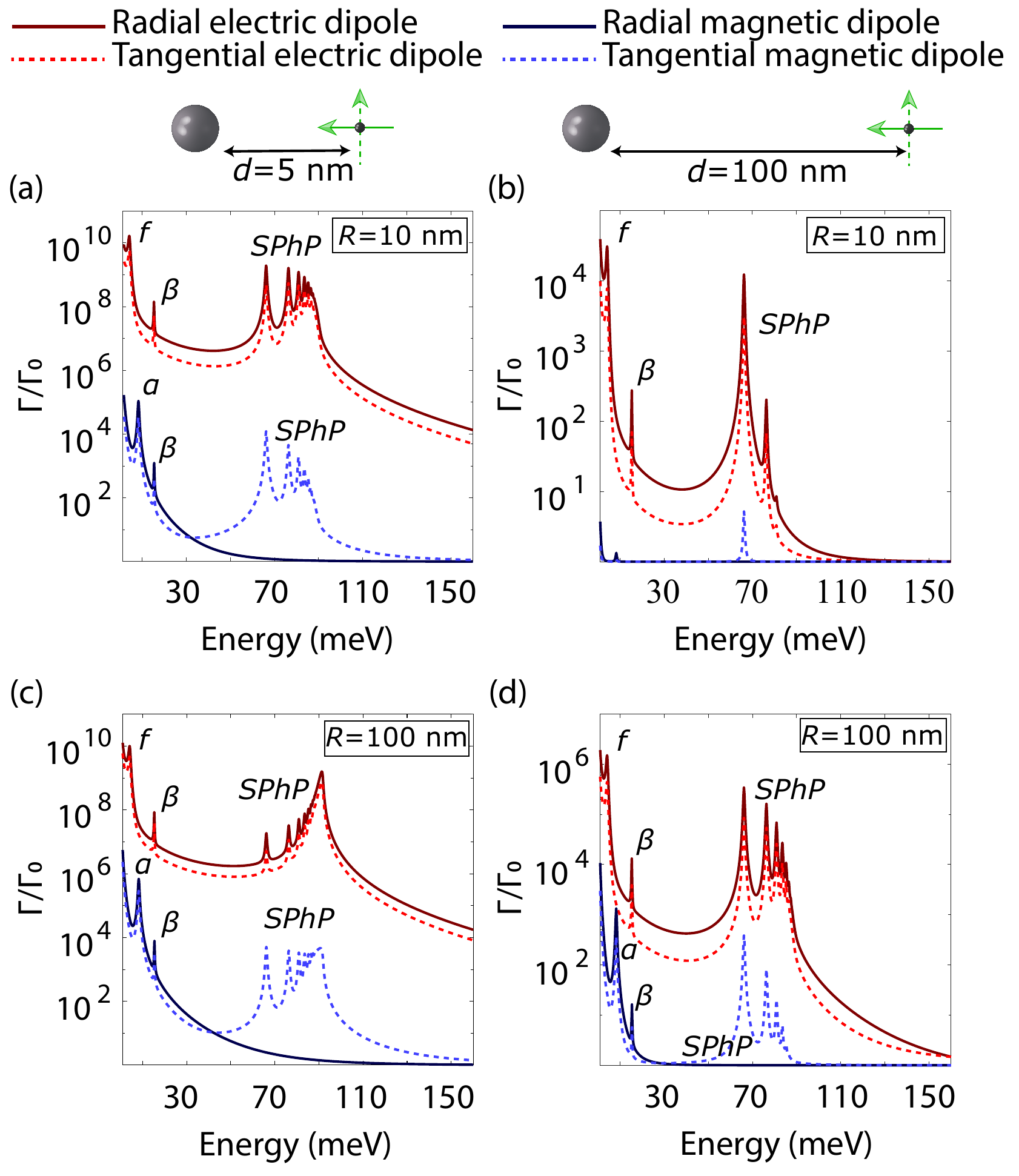}
\caption{PFs (log scale) for a QE with electric (red line) or magnetic
(blue line) dipole transitions, radially (solid lines) or tangentially
(dashed lines) oriented with respect to the surface of a Bi$_{2}$Se$_{3}$
nanosphere of radius $R$, as shown in the schematics, in the absence of
surface states.
(a) The QE is placed at a distance of $d = 5$\,nm, and $R=10$\,nm;
(b) $d=100$\,nm, and $R = 10$\,nm;
(c) $d = 5$\,nm, and $R = 100$\,nm;
(d) $d = 100$\,nm, and $R=100$\,nm.}
\label{fig5}
\end{figure}

Based on the Fano shape that characterizes the new topological
modes, we expect that the rate of excitation of QEs that have intrinsic
magnetic transitions in the region of THz, such as erbium ions
(Er$^{3+}$)~\cite{mikhaylovskiy_prl118}, can be drastically modified.
The interaction of TI NPs with QEs with electric transitions has been
investigated within the quasistatic regime, revealing the presence of
high electric PFs~\cite{thanopulos_ol47}. In view of our discussion above,
we now explore the complete EM solution for the interaction of QEs with
a Bi$_{2}$Se$_{3}$ nanosphere, taking into consideration all the
aforementioned phenomena (topological surface current, magnetic
modes, surface mode splitting). Because of the presence of magnetic
modes, we also consider magnetic dipoles, whose transitions are typically
many orders of magnitude slower. The electric (E) and magnetic (M) PFs
are calculated by the following relations~\cite{schmidt_oex13}
\begin{subequations}
\begin{align}\label{magnetic_purcell}
\frac{\Gamma_{E}^{\perp}}{\Gamma_{0}} &= 
1 - \frac{3}{2}
\mathrm{Re} \sum_{\ell = 1}^{\infty}
(2\ell + 1) \ell (\ell + 1)
a_{\ell} \Bigg[
\frac{h_{\ell}^{+} (x_{2})}{x_{2}}
\Bigg]^{2} \\
\frac{\Gamma_{E}^{\parallel}}{\Gamma_{0}} &= 
1 - \frac{3}{4} (2\ell + 1) \mathrm{Re}
\sum_{\ell = 1}^{\infty} \Bigg\{
a_{\ell} \bigg[
\frac{\xi_{\ell} (x_{2})}{x_{2}} \bigg]^{2} +
b_{\ell} h_{\ell}^{+2} (x_{2}) \Bigg\} \\
\frac{\Gamma_{M}^{\perp}}{\Gamma_{0}} &= 
1 - \frac{3}{2} \mathrm{Re} 
\sum_{\ell = 1}^{\infty}
(2\ell + 1) \ell (\ell + 1)
b_{\ell} \Bigg[
\frac{h_{\ell}^{+} (x_{2} )} {x_{2}} 
\Bigg]^{2} \\
\frac{\Gamma_{M}^{\parallel}}{\Gamma_{0}} &= 
1 - \frac{3}{4} 
(2\ell + 1) \mathrm{Re}
\sum_{\ell = 1}^{\infty} \Bigg\{
b_{\ell} \bigg[
\frac{\xi_{\ell} (x_{2})}{x_{2}} \bigg]^{2} +
a_{\ell} h_{\ell}^{+2} (x_{2}) \Bigg\}
,
\end{align}
\end{subequations}
where the decay rate of an electric (magnetic) dipole in the case of a
radial and tangential orientation, relatively to the NP surface (see
sketches in Fig.~\ref{fig5}), is denoted as $\Gamma_{E}^{\perp}$
$(\Gamma_{M}^{\perp})$, and $\Gamma_{E}^{\parallel}$
$(\Gamma_{M}^{\parallel})$, respectively. The decay rate in vacuum is
denoted as $\Gamma_{0}$.

To highlight the extent of the changes that an electrodynamic description
can bring to the QE--NP interaction,
we first calculate in Fig.~\ref{fig5}
the PFs for a Bi$_{2}$Se$_{3}$ nanosphere \emph{without} surface states,
considering cases with radii of $10$\,nm and $100$\,nm, at QE-to-NP surface
distances of $d = 5$\,nm and $d = 100$\,nm. Both near and far from the NP
surface, significantly enhanced spontaneous emission rates are observed,
as a result of the strong interaction between the QE and the NP eigenmodes.
Specifically, for the electric-dipole case, resonances occur for frequencies
corresponding to the plasmonic mode $f$, the $\beta$ bulk phonon polariton
mode, and the SPhP. Notably, these resonances do not shift in energy by
varying the NP size or the emitter distance, but they do change in
linewidth. Within the resonance region of the SPhP ($70-100$\,eV),
additional peaks emerge due to the higher-order
multipolar SPhPs, which do not manifest in the extinction spectra,
but their existence was demonstrated in Figs.~\ref{fig2}(b) and (c).
Concerning the magnetic dipole case, resonances linked to both the
$\alpha$ bulk phonon polariton at about $8$\,meV and the $\beta$ bulk
phonon polariton at $15$\,meV emerge, aligning with the corresponding
resonances of the dielectric function. Similarly, coupling occurs with
the SPhPs, albeit only when the dipole is oriented parallel to the NP
surface, showcasing coupling with analogous higher-order multipolar
modes. 
Our calculations show that the first SPhP mode is the main
contributor to the radiative enhancement of the Purcell effect, while
the other resonances correspond to nonradiative decay.
It is remarkable that the PF values for a magnetic dipole interacting
with the bulk $\alpha$ phonon mode are of the order of $10^{6}$, while
they still reach values of $10^{4}$ when interacting with the other
modes. The reason for such a strong light--matter interaction of
magnetic nature lies in the fact that the high-refractive-index
dielectric generates displacement current loops, inducing magnetic resonances.

\begin{figure}
\centering
\includegraphics[width=1.0\linewidth]{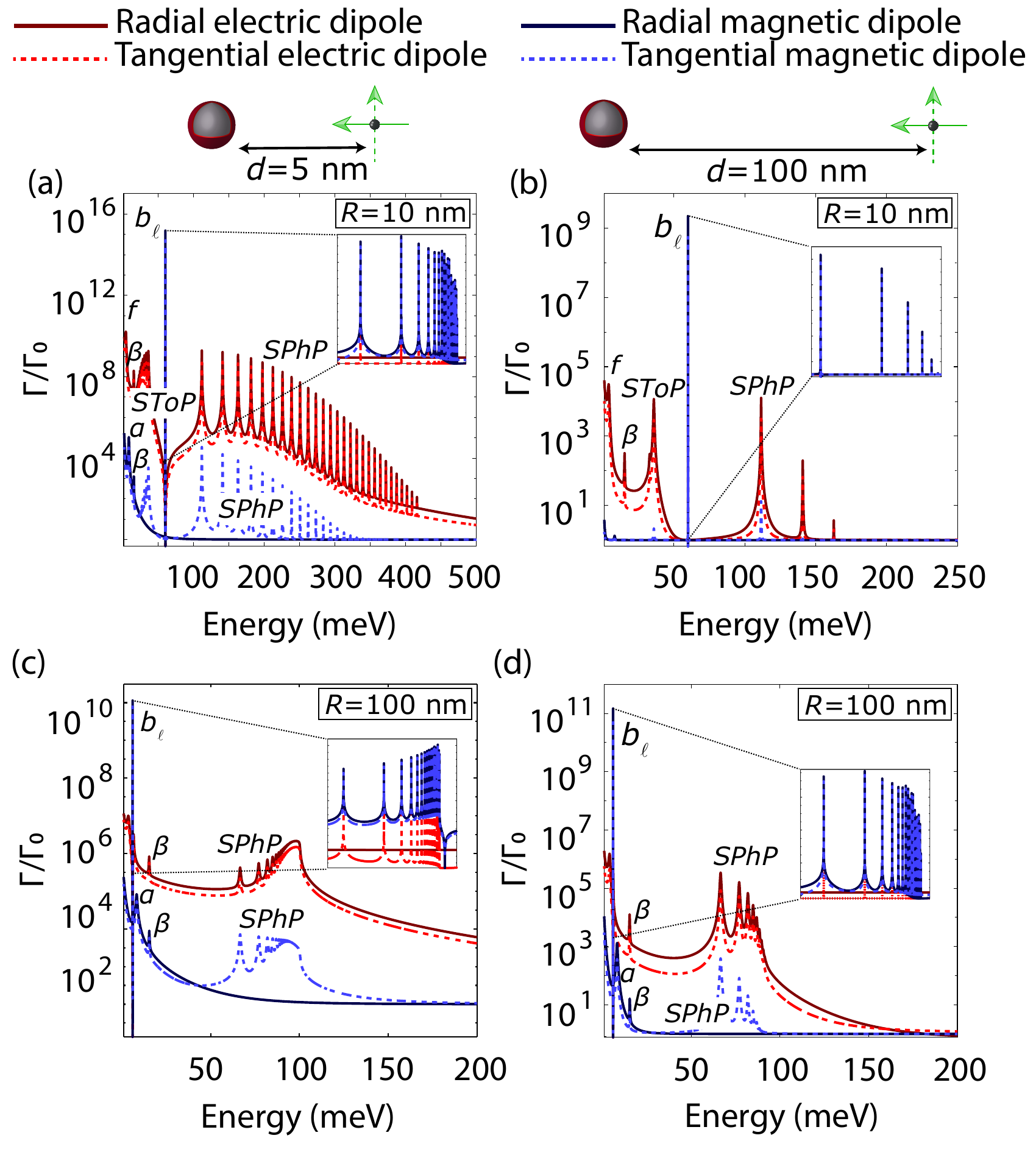}
\caption{Same as Fig.~\ref{fig5}, but with surface-states taken into
account. The insets zoom in at the energies of the $b_{\ell}$ resonances.}
\label{fig6}
\end{figure}

\begin{figure*}[ht]
\centering
\includegraphics[width=0.9\linewidth]{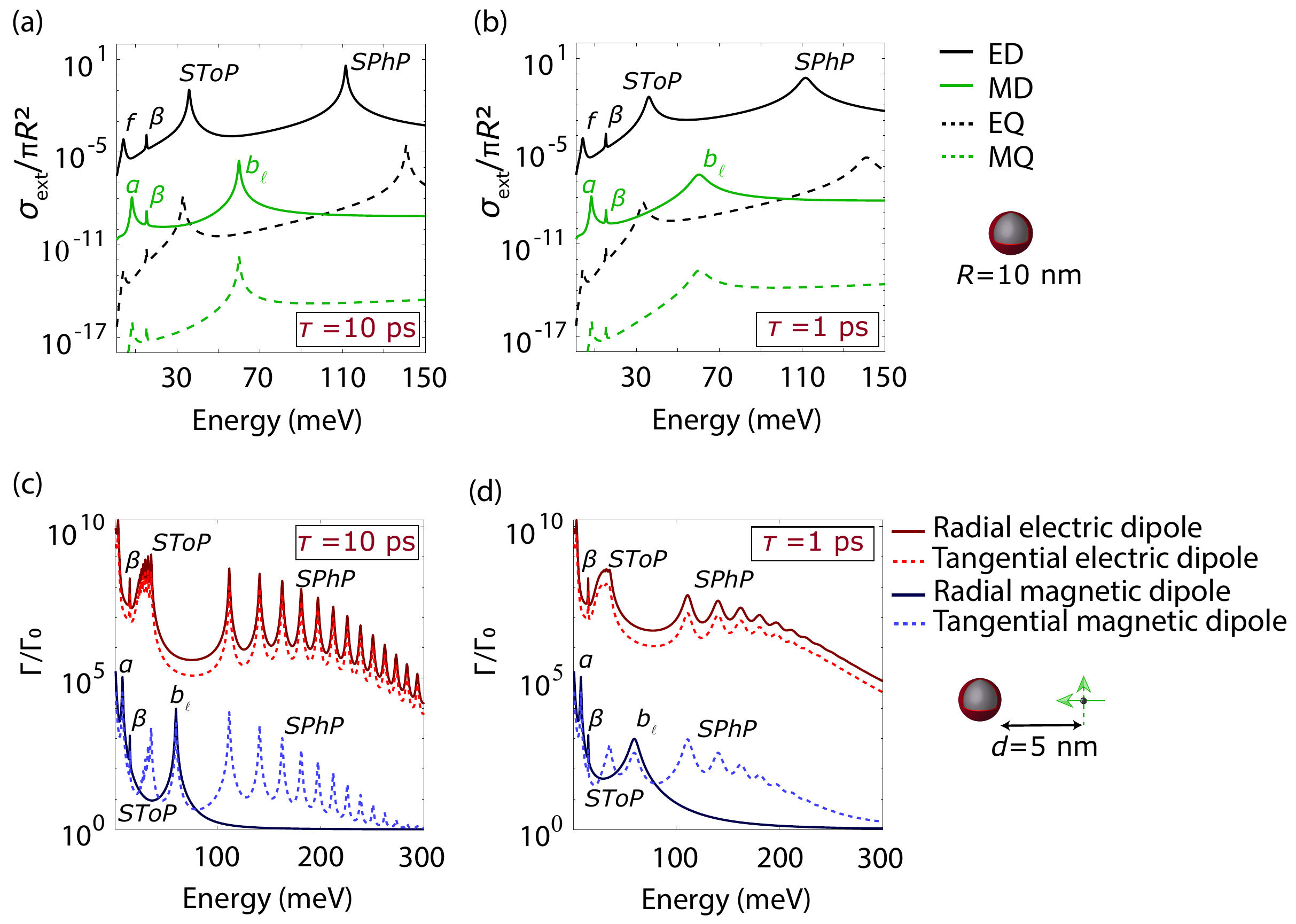}
\caption{(a) Decomposition of the normalized extinction cross section
of a Bi$_{2}$Se$_{3}$ nanosphere of radius $R = 10$\,nm into ED (solid
black line), EQ (dashed black line), MD (solid green line), and MQ
(dashed green line) contributions, considering a damping mechanism
with $\tau = 10$\,ps.
(b) Same as (a), for $\tau = 1$\,ps. 
(c) PFs of electric- (red lines) or magnetic-dipole (blue lines)
transitions, radially (solid lines) and tangentially (dashed lines)
oriented to the NP surface, for the NP of (a, b), at QE--NP distance
$d = 5$\,nm, and $\tau = 10$\,ps.
(d) Same as (c), for $\tau = 1$\,ps.}
\label{fig7}
\end{figure*}

Extending our investigation, we provide in Fig.~\ref{fig6}
the corresponding PFs when surface states are included
[for a comparison with the quasistatic approach, see Fig.~\ref{figS4}(b)], and
analyze further the influence of DPPs on the QE--NP coupling.
The impact of surface states dramatically alters the spectrum, as
expected for surface polaritons. The resilience of the plasmonic $f$
mode and the $\beta$ bulk phonon polariton is still obvious, but the
impact of surface states becomes more evident in the enhanced PF values
observed at the frequencies of the DPPs.
Specifically, DPPs couple efficiently with the electric dipole,
leading to pronounced SToP-related PF enhancement, while their coupling
with the magnetic dipole leads to PF enhancement due to the topological
magnetic modes ($b_l$), manifesting as the needle-like resonances shown
in the insets. Comparing with Fig.~\ref{fig5}, shifting of the
resonances of the spontaneous emission rate is also noticeable, as the
SPhP splitting emerges for $R = 10$\,nm.
Furthermore, in the case where the QE is positioned at a distance
of $5$\,nm [Figs.~\ref{fig6}(a) and (c)], the Purcell enhancement factors
exhibit a colossal increase, while, at the same time, the emitter couples
with many higher-order multipolar modes. These results imply that the QE
interacts strongly with the conducting surface of the nanosphere,
specifically with the surface charge density which is coupled with
the SPhP, giving rise to DPPs. The strong interaction arising from
intense oscillations of the induced surface current is confirmed by
the zoomed-in frequency regions surrounding the induced magnetic
eigenmodes. It is also noteworthy that these high PFs survive even
at significant distances from the NP surfaces, as shown in
Figs.~\ref{fig6}(b) and (d).

The interaction of an electric-dipole QE with the DPP excited at the
surface of the TI results in a finite PF, as expected. The mechanism
that prevents higher values is the phonon-induced damping rate introduced
into the dielectric function. However, the interaction of magnetic-dipole
QEs with the $b_{\ell}$ magnetic modes yields
unphysically high magnetic PFs. As we mentioned earlier, the
origin of delta-like resonances in either extinction or PF can be
attributed to the absence of a finite lifetime for the surface states
in Eq.~\eqref{delta}.
To address this issue, we introduce in Fig.~\ref{fig7} damping rates
for the surface states by hand, to account for the finite lifetimes
of the excited Dirac electrons. More specifically, we introduce an
imaginary term $\mathrm{i} \Gamma$ into the denominators of 
Eq.~\eqref{delta}, which is now modified to
\begin{align}\label{delta2}
\delta_{R} (\omega) =
\frac{e^{2}}{6 \pi \varepsilon_{0} R}
\Bigg(
\frac{1}{2 \Delta\mathscr{E} - \hbar\omega -\mathrm{i} \Gamma} +
\frac{1}{2 \Delta\mathscr{E} + \hbar\omega - \mathrm{i} \Gamma}
\Bigg)
.
\end{align}
In this expression, we use for $\Gamma$ two different values
based on the lifetimes of DPPs~\cite{sobota_prl108,russmann_jpcs128,
hale_aplphot8,pogna_natcom12}, although we anticipate that these values,
which correspond to calculations or measurements of surface states in
large systems and with several co-acting damping mechanisms, might
heavily underestimate the lifetime of the transitions between
quantized surface states. It is also worth keeping in mind that, since
the transitions under study only concern one or two surface electrons
per Brillouin zone, the corresponding recombination channels are rather
limited, and the excited states are expected to be quite long-lived.
Focusing on a sphere with $R = 10$\,nm, we initially incorporate a
$\Gamma = 0.1$\,THz, corresponding to lifetime $\tau = 10$\,ps;
subsequently we shift to a---possibly extreme---$\Gamma = 1$\,THz,
corresponding to $\tau=1$\,ps. For both lifetimes, the extinction
cross sections in Figs.~\ref{fig7}(a) and (b) demonstrate that
both the topological electric mode and the resonance splitting due to
the interaction of the SPhP and the DPP persist. We thus conclude from
Fig.~\ref{fig7}(b) that, even for high damping rates of the excited 
surface states, the NP sustains its topological optical modes. We 
calculated similar broadening, with the main resonances still surviving,
for an intermediate radius $R = 50$\,nm, plotted in Fig.~\ref{figS3}.

Subsequently, we calculate in Figs.~\ref{fig7}(c) and (d) the corresponding
PFs, for a QE positioned at distance $d = 5$\,nm from the NP surface.
Interestingly, we observe comparable resonances and footprints as in the
infinite-lifetime case. As expected, the topological magnetic Purcell
enhancement now attains a finite---yet significantly high---value of
$\sim 10^{4}$, while the topological electric PF assumes values of
$\sim 2.0 \times 10^{8}$. For larger spheres, such as $R= 50$\,nm
at QE-to-NP surface distances of $d = 5$\,nm, the magnetic (electric) 
PF originating from the excitation of DPPs is still considerable, as
high as $\sim 1.0 \times 10^{5} (\sim 3.0 \times 10^{8})$ considering
the same lifetimes as in Figs.~\ref{fig7}(c) and (d) (see Fig.~\ref{figS3}).
The PFs are higher for this larger-dimension case, because the topological
surface currents circulate on a larger NP surface. These results suggest
that even for large damping rates,
the TI NP still manifests magnetic modes due to the existence
of DPPs.

\section{Conclusions}
We employed fully electrodynamic analytical calculations based on
Mie theory, to demonstrate the emergence of topological magnetic modes
resulting from the interaction of THz radiation with Bi$_{2}$Se$_{3}$
spherical TIs. In particular, we observed the excitation of previously
unreported magnetic modes, induced by topological surface currents
that are related to transitions between discrete energy levels in the
Dirac cone, generating currents along the conductive surface of the NPs.
For specific NP sizes, these topological surface currents can even lead to
striking mode splitting in absorption and scattering spectra, as a result
of the self-hybridization between DPPs and SPhPs. These modes,
distinguished by their Fano-like
shape, produce robust coupling effects over a wide window of energies,
for QEs at a broad range of distances from the NP. Additionally, we
identified strong light--matter interaction of magnetic nature, since
the high-refractive-index dielectric generates displacement current
loops in the bulk, stimulating substantial Mie-type magnetic response.
As a result, we calculated remarkable magnetic and electric Purcell
enhancement over a wide energy window from a few THz to a few hundreds
of THz. This response arises from the synergy between topological
surface plasmons and bulk NP modes. The magnetic PFs obtained through
the coupling of QEs with bulk phonon polaritons are notably high,
reaching values of up to $\sim 7\times 10^{5}$. Simultaneously, the
magnetic PFs associated with DPPs remain significantly large
($\sim 10^{4}$ to $10^{5}$) even when realistic lifetimes for the
surface states are considered. Bi$_{2}$Se$_{3}$ nanospheres demonstrate
thus a rich optical response that pertains to both the bulk and the
topologically protected surface states, enabling the excitation of DPPs,
and are positioned as promising candidates for facilitating strong
light--matter interactions in nanophotonics and quantum optics. Our
results have the potential to trigger further research for QEs with
magnetic transitions in the THz, or for the acceleration of otherwise
forbidden atomic transitions.

\vspace{1.0cm}

\section*{Acknowledgments}
N.~A.~M. is a VILLUM Investigator supported by VILLUM Fonden
(grant No. 16498).
The Center for Polariton-driven Light--Matter Interactions (POLIMA) is sponsored by the Danish National Research Foundation (Project No.~DNRF165).

\newpage
\appendix
\renewcommand{\theequation}{S.\arabic{equation}}
\setcounter{equation}{0}
\renewcommand{\thefigure}{S.\arabic{figure}}
\setcounter{figure}{0}

\section{Full EM calculations vs quasistatic regime}

\begin{figure}[h]
\centering
\includegraphics[width=1.0\linewidth]{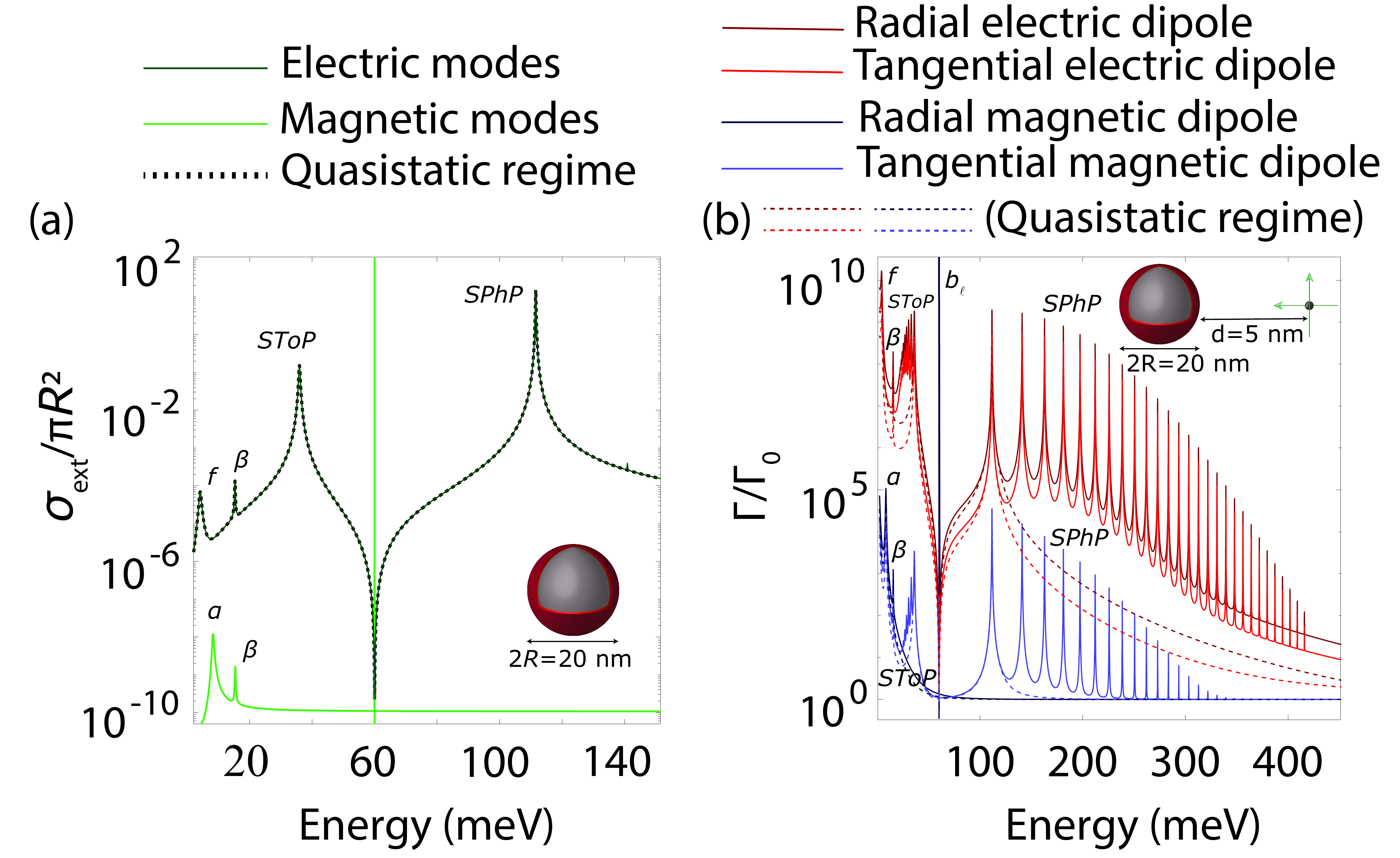}
\caption{(a) Normalized extinciton cross section (log scale) of a
Bi$_{2}$Se$_{3}$ nanosphere of radius $R = 10$\,nm in the cases of the
quasistatic approach (dashed lines) and in the case of the full 
electrodynamic solution (solid lines).
(b) Purcell factors (log scale) with electric (red line) or magnetic
(blue line) dipole transitions, radially (dark color) and tangentially
(light color) oriented to the NP surface, with radius same as (a) and
emitter--nanoparticle distance $d = 5$\,nm, for the quasistatic regime
(dashed lines) and for all contributions from the multipolar expansion
(solid lines).}
\label{figS4}
\end{figure}

In order to elucidate the significance of the full electromagnetic solution, which includes all multipolar contributions (solid lines), and to emphasize the difference from the quasistatic approximation, i.e., only the electirc dipole modes (dashed lines), we provide Fig.~\ref{figS4}. It should be further noted that the solid lines, associated with Mie resonances and the new topological magnetic modes, are not discernible with the quasistatic approximation.

\vspace{1.5cm}

\vspace{1.0cm}

\section{Extinction of the STOP mode}

\begin{figure}[ht]
\centering
\includegraphics[width=1.0\linewidth]{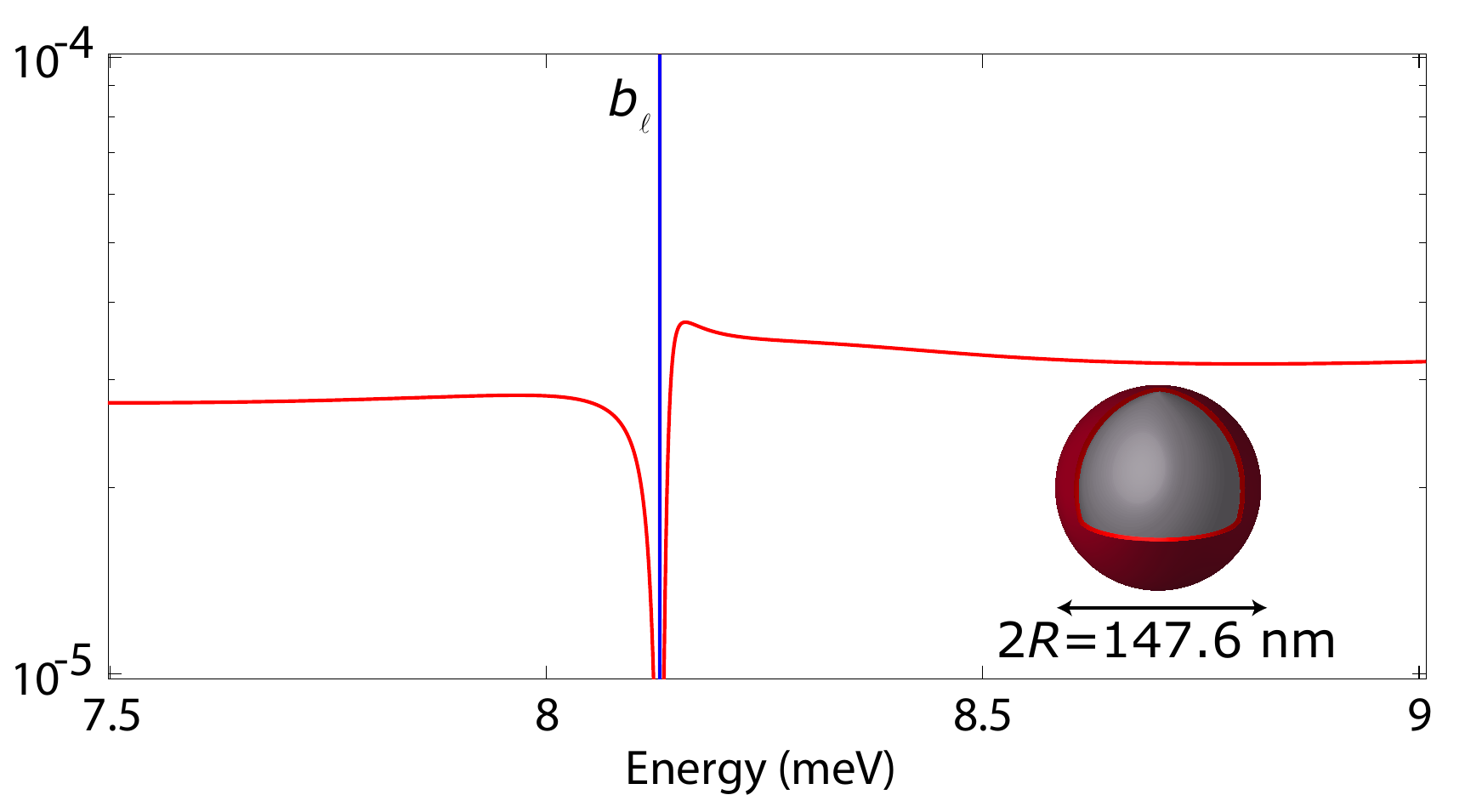}
\caption{Normalized absorption (red lines) and scattering (blue
lines) cross sections (log scale) of a Bi$_2$Se$_3$ nanosphere in which
transitions between discrete levels at the surface states have
been taken into account, for $R = 73.8$\,nm.}
\label{figS2}
\end{figure}

In Fig.~\ref{figS2}, we present the optical response of a 
Bi$_{2}$Se$_{3}$ nanosphere with a radius of $R = 73.8$\,nm,
focusing on the energy region where the excitation of surface
states occurs, aiming to distinguish between the modes. The
specific value for the radius represents the unique condition
at which the surface topological optical (SToP) mode disappears.
Nevertheless, the induced magnetic modes $b_{\ell}$ remain, as
indicated by the graph.

\vspace{1.5cm}

\vspace{1.0cm}

\section{Multipolar decomposition}

\begin{figure}[hb]
\centering
\includegraphics[width=1.0\linewidth]{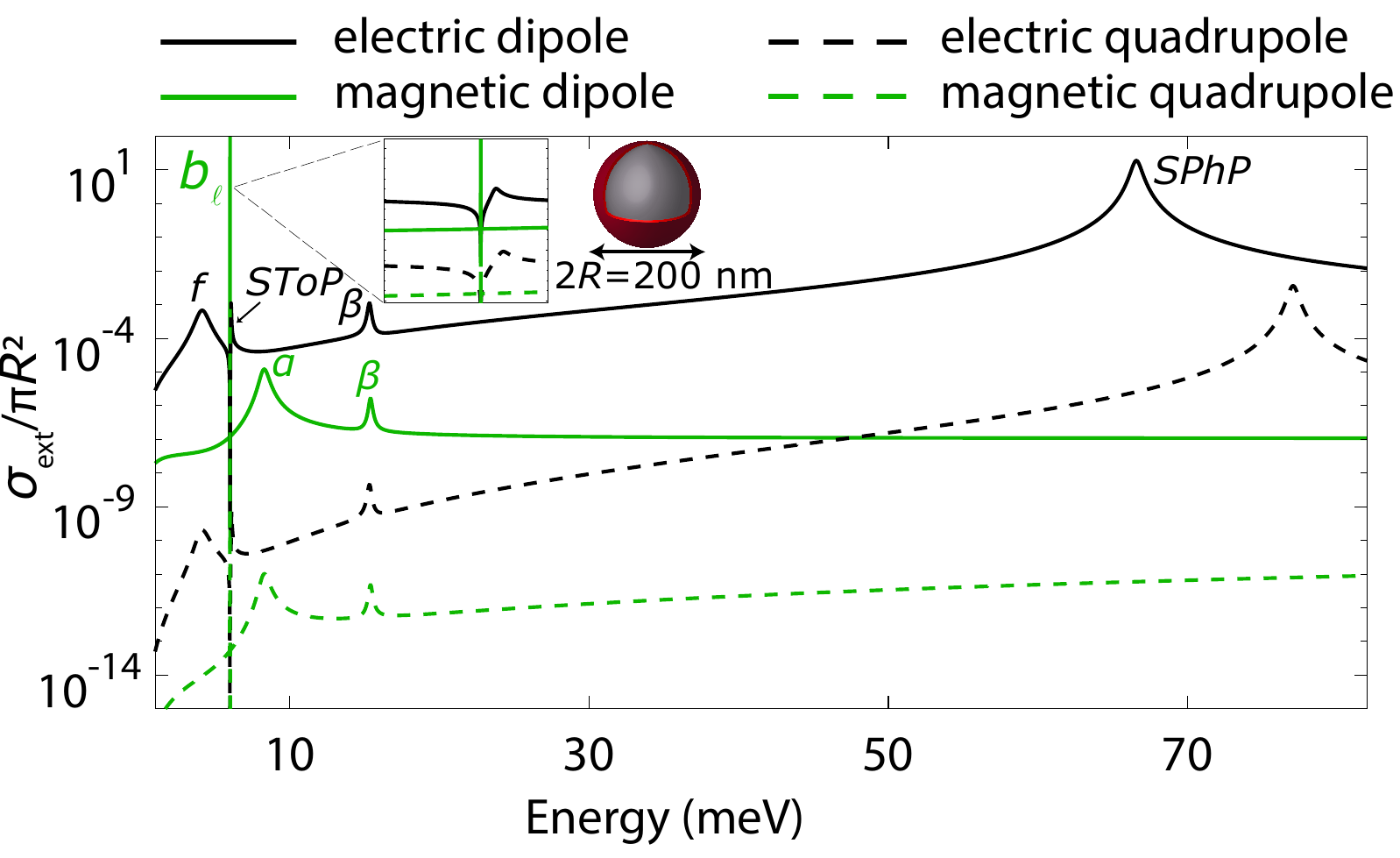}
\caption{Multipolar decomposition via Mie coefficients of electric dipolar
(black solid line), magnetic dipolar (dashed black line), electric 
quadrupolar (green line), and magnetic quadrupolar (dashed green line) 
contributions to the normalized extinction cross section (log scale) for
a spherical topological insulator with $R = 100$\,nm.}
\label{figS1}
\end{figure}

We extend the analysis of the optical response of a spherical topological
insulator in Fig.~\ref{figS1}, wherein a detailed representation of the
contributions from the multipolar expansion, for a nanosphere with a
radius of $R = 100$\,nm is depicted.

\vspace{1.0cm}

\newpage

\section{Role of damping}

\begin{figure}[h]
\centering
\includegraphics[width=1.0\linewidth]{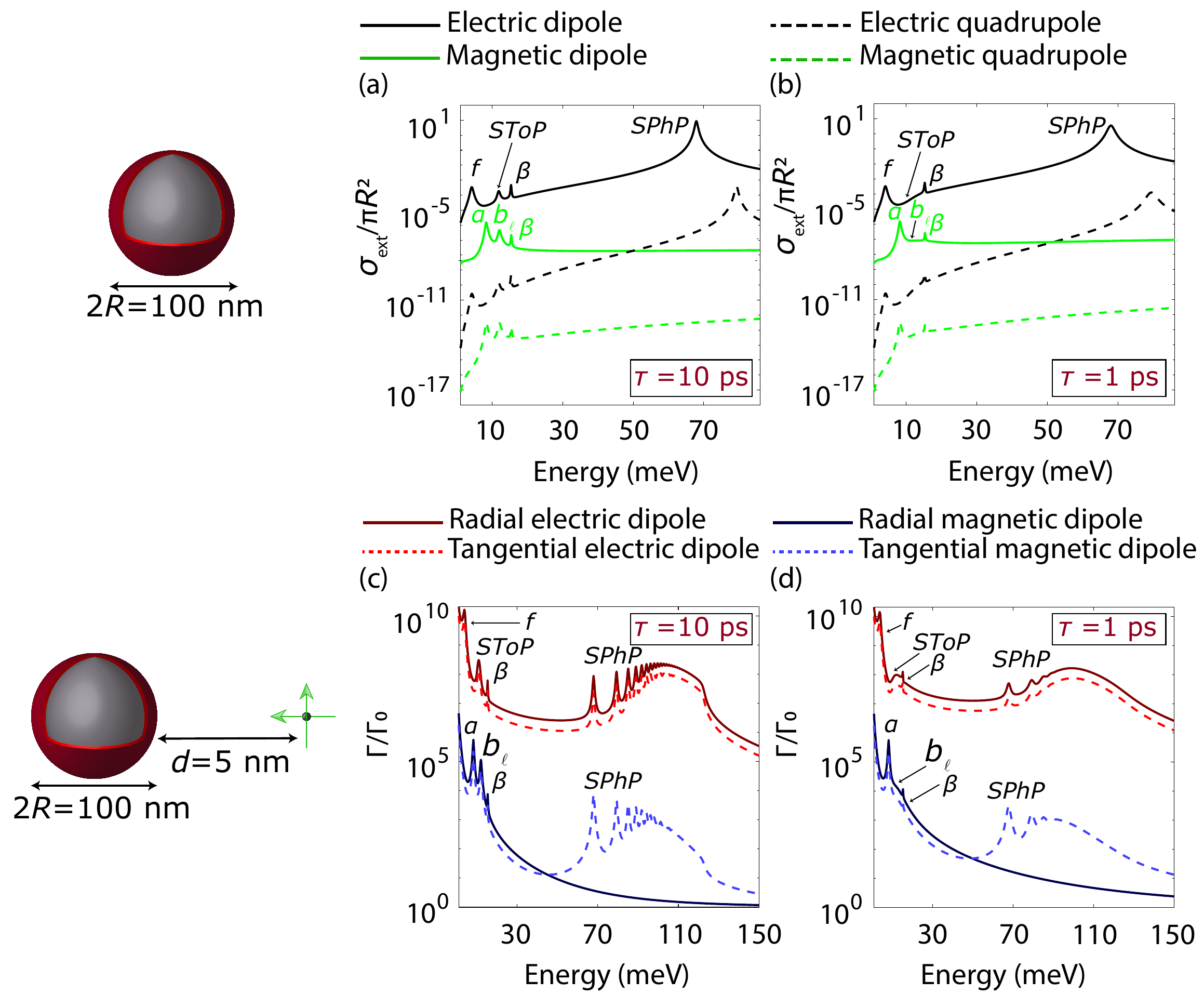}
\caption{(a, b) Decomposition of the normalized extinction cross
section (log scale) of a Bi$_{2}$Se$_{3}$ nanosphere of radius
$R = 50$\,nm into electric dipolar (solid black line), electric
quadrupolar (dashed black line), magnetic dipolar (solid green line),
and magnetic quadrupolar (dashed green line) contributions, considering
a damping mechanism with
(a) $\tau=10$\,ps and
(b) $\tau=1$\,ps.
(c, d) Purcell factors with electric- (red line) or magnetic-dipole
(blue line) transitions, radially (dark color) and tangentially (light
color) oriented to the nanoparticle surface, with radius same as in
(a, b), emitter--nanoparticle distance $d = 5$\,nm, and
(c) $\tau = 10$\,ps or
(d) $\tau = 1$\,ps.}
\label{figS3}
\end{figure}

For confirmation of the persistence of Dirac plasmon polaritons
when introducing a limited lifetime of excited states, we study in
Fig.~\ref{figS3} a nanoparticle with radius $R = 50$\,nm. The
multipolar decomposition of the extinction cross section is shown in
Figs.~\ref{figS3}(a) and (b), while the corresponding Purcell factors
are plotted in Figs.~\ref{figS3}(c) and (d), for emitter--nanoparticle
distance of $d = 5$\,nm. We explore two different lifetimes,
$\tau=1$\,ps and $\tau=10$\,ps.

\newpage

\bibliography{references}

\end{document}